\documentclass[twocolumn,showpacs,preprintnumbers,amsmath,amssymb,aps]{revtex4-1}
\usepackage{graphicx}
\usepackage{dcolumn}
\usepackage{bm}
\usepackage[unicode=true,colorlinks=true,citecolor=blue,urlcolor=blue]{hyperref}
\usepackage{epstopdf}
\usepackage[T1]{fontenc} % special letters
\usepackage{lmodern}
\usepackage{braket}
\usepackage{color}
\definecolor{color}{rgb}{0.11,0.45,0.02}

\begin{document}

\title{Dynamic polarization of electron spins in indirect band gap (In,Al)As/AlAs quantum dots in weak magnetic field: experiment and theory}

\author{T.~S.~Shamirzaev$^{1,2}$, A.~V.~Shumilin$^{3}$, D.~S.~Smirnov$^{3,4}$, J.~Rautert$^5$, D.~R.~Yakovlev$^{5,3}$, and M.~Bayer$^{5,3}$}

 \affiliation{
 $^1$Rzhanov Institute of Semiconductor Physics, Siberian Branch of the Russian Academy of Sciences, 630090 Novosibirsk, Russia \\
 $^2$Ural Federal University, 620002 Yekaterinburg, Russia\\
 $^3$Ioffe Institute, Russian Academy of Sciences, 194021 St.Petersburg, Russia\\
 $^4$Spin Optics Laboratory, St. Petersburg State University, 198504 St. Petersburg, Russia\\
 $^5$Experimentelle Physik 2, Technische Universit\"at Dortmund, 44227 Dortmund, Germany
}

\begin{abstract}
A novel spin orientation mechanism --- dynamic electron spin
polarization has been recently suggested in Phys. Rev. Lett.
\textbf{125}, 156801 (2020). It takes place for unpolarized optical
excitation in weak magnetic fields of the order of a few millitesla.
In this paper we demonstrate experimentally and theoretically that
the dynamic electron spin polarization degree changes sign as a
function of time, strength of the applied magnetic field and its
direction. The studies are performed on indirect band-gap
(In,Al)As/AlAs quantum dots and their results are explained in the
framework of a theoretical model developed for our experimental
setting.
\end{abstract}

\maketitle

\section{Introduction}
\label{sec:intro}

The manipulation of the spin degree of freedom in semiconductor
nanostructures is interesting  from the viewpoints of both basic
physics~\cite{Dyakonov,Glazov} and  potential
applications~\cite{Dyakonov1,Fert}. Generation of electron and hole
spin polarization can be achieved in several ways. The main
approaches to spin orientation are: optical spin
orientation~\cite{OO_book} and thermal spin polarization in a
magnetic
field~\cite{zeeman1897effect,Shamirzaev60,Shamirzaev96,Shamirzaev94}.
The first one is based on the transfer of the angular momentum from
circularly polarized photons to electrons through the spin-orbit
interaction. The second approach requires a lowering of the lattice
temperature, so that the thermal energy becomes smaller than the
electron Zeeman splitting. It can be enhanced in nonequilibrium
conditions~\cite{Ivchenko60}.

Recently we have suggested another approach to spin polarization:
the dynamic electron spin polarization~\cite{Smirnov125}. In
contrast to the optical spin orientation, the proposed mechanism
does not require circular polarization of the optical excitation. In
contrast to the thermal spin polarization, we consider  weak
magnetic fields, for which the electron Zeeman splitting is much
smaller than the thermal energy. This mechanism was approved
experimentally for (In,Al)As/AlAs quantum dots
(QDs)~\cite{Smirnov125}.

In this paper, we extend these studies and present a deeper
experimental and theoretical investigation of the dynamic electron
spin polarization in indirect band gap (In,Al)As/AlAs QDs with type
I band alignment. Namely, the dynamics of spin polarization in magnetic
fields with different strength and orientation are studied. We
find experimentally that the electron spin polarization degree
changes sign with increasing delay time after pulsed excitation with unpolarized light or
with the strength and orientation of the weak magnetic field. These
experimental results agree with the developed microscopic theory based
on the hyperfine interaction with nuclear spin fluctuations and the
exchange interaction between electron and hole in an exciton confined in a QD.

The paper is organized as follows. In Sec.~\ref{sec:1} the studied sample
and the applied experimental techniques are described. In
Sec.~\ref{sec:2} we present the experimental results, before in
Sec.~\ref{sec:theory} we develop a theoretical model and describe
general results obtained with the model. In Sec.~\ref{sec:compare} we compare
the experiment and theory.

\section{Sample and experimental setup}
\label{sec:1}

The studied self-assembled (In,Al)As QDs embedded in an AlAs matrix
were grown by molecular-beam epitaxy on a semi-insulating
$(001)$-oriented GaAs substrate with a 400-nm-thick GaAs buffer
layer~\cite{Shamirzaev78}. The structure contains one QD layer
sandwiched between two 70-nm-thick AlAs layers. The nominal amount
of  deposited  InAs is about $2.5$ monolayers. The QDs were formed
at the temperature 520$^{\circ}$C with a growth interruption time
of 20~s. The top AlAs barrier is protected from oxidation by a
$20$-nm-thick GaAs cap layer. From the growth conditions and model
calculations, we conclude that the average QD composition is
In$_{0.75}$Al$_{0.25}$As~\cite{Shamirzaev78}. The size and density of
the lens-shaped QDs were measured by transmission electron
microscopy, yielding an average diameter of 12~nm and a density of
about 2$\times$10$^{10}$ dots per cm$^2$. The relatively low QD
density prevents carrier redistribution between the
QDs~\cite{ShamirzaevAPL97,ShamirzaevSST}.

The sample was placed in a liquid helium bath cryostat. The
temperature for all experiments was fixed at $T = 1.8$~K. Low
magnetic fields in the millitesla range were generated by an
electromagnet. The angle $\theta$ between the magnetic field
direction and the QD growth axis (z axis) was varied between
0$^{\circ}$ (Faraday geometry) and 90$^{\circ}$ (Voigt geometry).
For measurement of angular dependencies we fixed the magnetic field
direction and rotated the sample. The emission was collected either
in the direction along the field axis in Faraday geometry for
$0^{\circ} \leq \theta \leq 45^{\circ}$ or perpendicular to the
field axis in Voigt geometry for $45^{\circ} \leq \theta \leq
90^{\circ}$.

The photoluminescence (PL) was excited nonresonantly with the laser
photon energy exceeding the direct band gap of the AlAs matrix,
which is equal to 3.099~eV~\cite{Vurgaftman}. We used the third
harmonic of a Q-switched Nd:YVO$_4$ pulsed laser with a photon
energy of $3.49$~eV, a pulse duration of $5$~ns and a repetition rate of
1~kHz~\cite{Shamirzaev84}. The exciting light was linearly polarized,
and we checked that the direction of the polarization does not
affect the presented results, as expected for strongly
nonresonant excitation. The PL was dispersed by a $0.5$-m
monochromator. For the time-resolved and time-integrated PL
measurements we used a gated charge-coupled-device (CCD) camera,
synchronized with the laser via an external trigger signal. The time
delay between the pump pulse and the start of PL recording,
$t_{\text{d}}$, was varied from zero up to $1$~ms. The duration
of PL recording, i.e., the gate window $t_{\text{g}}$ was
varied from $1$~ns to $1$~ms. The signal intensity and the time
resolution of the setup depend on $t_{\text{d}}$ and $t_{\text{g}}$.
The highest time resolution of the detection system is $1$~ns.

The circularly polarized components of the emission were selected by
combining Glan-Thompson prisms and quarter-wave plates. The circular
polarization degree of the PL is given by $P_c
=(I_{+}-I_{-})/(I_{+}+ I_{-})$, where $ I_{+}$ and $I_{-}$ are the
intensities of the $\sigma^{+}$ and $\sigma^{-}$ polarized PL
components, respectively. To determine the sign of $P_c$, we
performed a control measurement on a diluted magnetic
semiconductor structure with (Zn,Mn)Se/(Zn,Be)Se quantum wells. For
this structure, it is known that $P_c>0$ for $B_z>0$ in Faraday
geometry~\cite{Keller}.

\section{Experimental results}
\label{sec:2}

The dispersion in QD size, shape, and composition within the
ensemble leads to the formation of (In,Al)As/AlAs QDs with different
band structure, as shown in Fig.~\ref{fig1}(a). The electron ground
state changes from the $\Gamma$ to the X valley with decrease of the
dot size, while the heavy-hole (hh) ground state remains at the
$\Gamma$ point. This corresponds to a change from a direct to an
indirect band gap. On the other hand, the type-I band alignment is
preserved, that is, in both cases, electron and hole are spatially
confined within the (In,Al)As
QDs~\cite{Shamirzaev78,Shamirzaev84,ShamirzaevAPL92}. Note that the
change of electron ground state from a direct to an indirect band
with decreasing QDs size is not unique for (In,Al)As/AlAs QDs, it
was demonstrated for various semiconductor
heterostructures~\cite{Cedric1,Cedric2,Apex,Shamirzaev45,Abramkin53,Pistol,Shamirzaev97}.

Recently, we demonstrated that the coexistence of (In,Al)As/AlAs QDs
with direct and indirect band gaps in the ensemble is evidenced
by the spectral dependence of the exciton recombination
time~\cite{Shamirzaev78,Shamirzaev84,ShamirzaevAPL92}. In direct
QDs, the excitons recombine within a few
nanoseconds~\cite{Rautert100}. On the contrary, the indirect QDs are
characterized by long decay times due to their small exciton
oscillator strength. Here, we use accordingly time-resolved photoluminescence to
identify indirect band-gap QDs.

\subsection{Time-resolved photoluminescence}

\begin{figure}[]
\includegraphics* [width=7 cm]{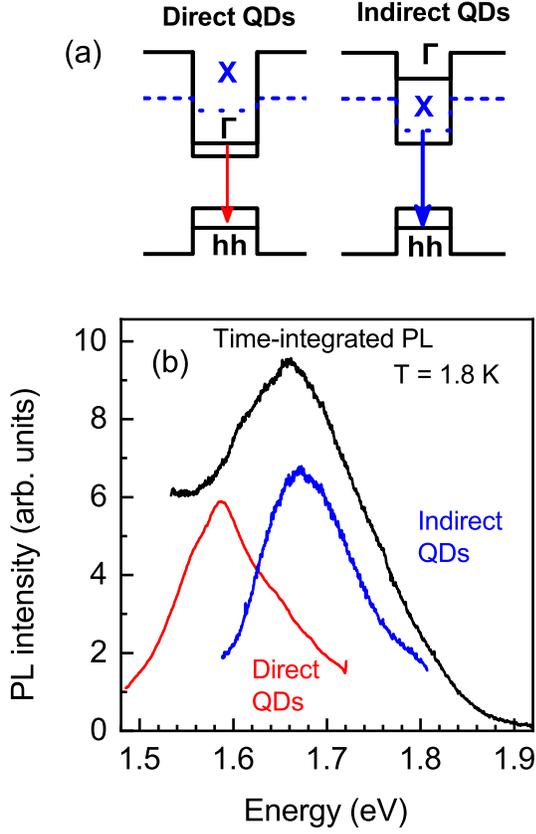}
\caption{\label{fig1} (a) Band diagrams of (In,Al)As/AlAs QDs with
direct and indirect band structure. (b)
Photoluminescence spectra of (In,Al)As/AlAs QDs measured for
non-resonant excitation: time-integrated (black line), time-resolved
for $t_{\text{d}}=1$~ns and $t_{\text{g}}=4$~ns (red), as well as for
$t_{\text{d}}=1~\mu$s and $t_{\text{g}}=5~\mu$s (blue).}
\end{figure}

Photoluminescence spectra of an (In,Al)As/AlAs QD ensemble measured
for nonresonant excitation are shown in Fig.~\ref{fig1}(b). The
time-integrated spectrum (black line) has a maximum at 1.67~eV and
extends from 1.50~eV (at lower energies the PL is related to the
GaAs substrate) to 1.90~eV, with a full width at half maximum (FWHM)
of about 190~meV. The large width of the emission band is due to the
dispersion of the QD parameters, since the exciton energy depends on
QD size, shape and composition~\cite{Shamirzaev78}. The PL band is
contributed by the emission of direct and indirect QDs, as becomes
evident from the time-resolved PL spectra. For the spectrum measured
immediately after the laser pulse ($t_{\text{d}}=1$~ns and
$t_{\text{g}}=4$~ns), the PL band (red line) has its maximum at
1.59~eV and a FWHM of 100~meV. For longer delays
($t_{\text{d}}=1~\mu$s and
 $t_{\text{g}}=5~\mu$s), the emission maximum shifts to 1.68~eV
and the band broadens to 180~meV (blue line), rather similar to the
time-integrated PL spectrum. We recently demonstrated that after
photoexcitation in the AlAs barriers electrons and holes are
captured in QDs within several picoseconds, and the capture
probability does not depend on the QD size and
composition~\cite{ShamirzaevNT}. Therefore, all QDs in the ensemble
(direct and indirect) become equally populated shortly after the
excitation pulse.

Thus, the exciton recombination dynamics is fast for direct QDs
emitting mainly in the spectral range of $1.50-1.70$~eV and slow for
the indirect QDs emitting in the range of $1.60-1.90$~eV. The
emission of the direct and indirect QDs overlaps in the range of
$1.60-1.70$~eV.

\begin{figure}[]
\includegraphics* [width=7cm]{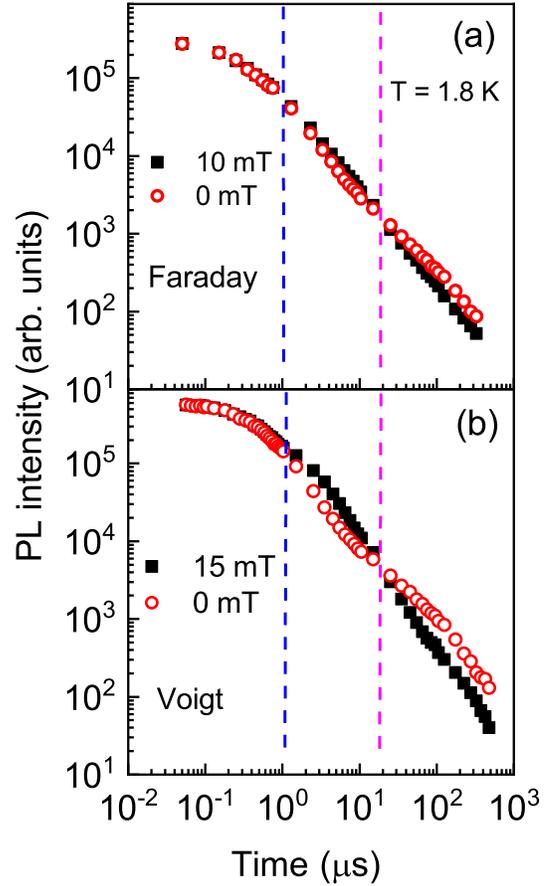}
\caption{\label{fig2} Dynamics of  unpolarized PL for indirect
excitons at the energy of $1.67$~eV in zero (open red
 circles) and weak magnetic fields (filled black squares)
for (a) Faraday and (b) Voigt geometry.  Dashed vertical lines
show delay times after excitation pulse $t_{\text{d1}}$ (blue) and
$t_{\text{d2}}$ (magenta). Note the bilogarithmic scale.}
\end{figure}

The dynamics of the unpolarized PL (sum of the $\sigma^{+}$ and
$\sigma^{-}$ polarized PL components) measured for indirect QDs at
1.67~eV in zero and in weak longitudinal and transverse magnetic
fields are shown in Figs.~\ref{fig2}(a) and \ref{fig2}(b),
respectively. The data are plotted in bilogarithmic scale, which is
convenient for presenting the dynamics across a wide range of decay
times and PL intensities. The decay curves are non-exponential. Such
a dynamics results from superposition of multiple monoexponential
decays of single QDs~\cite{Rautert100} with different decay times
varying with size, shape, and composition of the
QDs~\cite{Shamirzaev84}.

\begin{figure}[]
\includegraphics* [width=8.6cm]{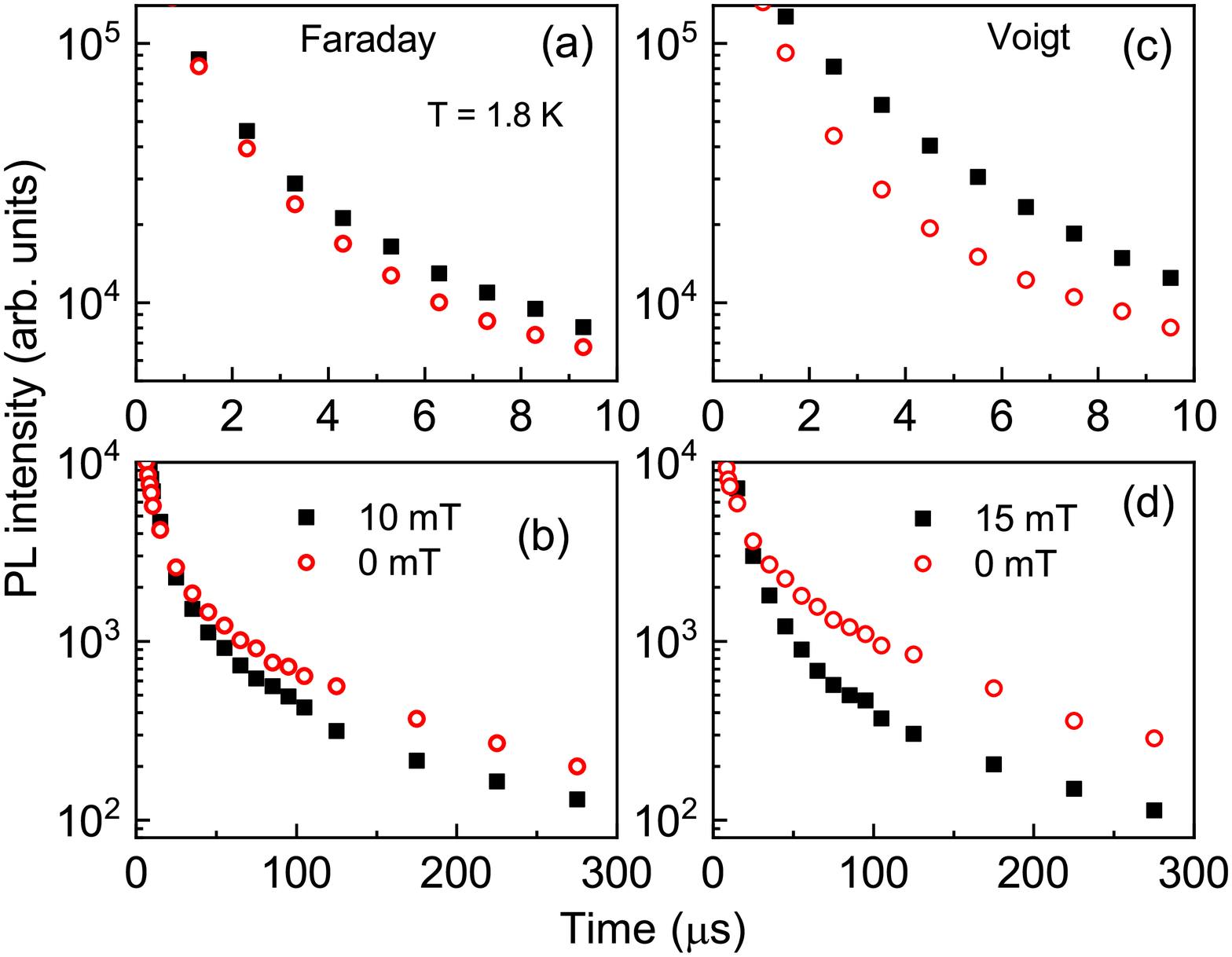}
\caption{\label{fig3} Dynamics of unpolarized PL for indirect
excitons at the energy of $1.67$~eV in zero (open red
 circles) and weak magnetic fields (filled black squares)
selected for different ranges of delay times after the excitation pulse:
(a) $0-t_{\text{d2}}$, (b) $t_{\text{d2}}-t_{\text{dmax}}$ for
Faraday geometry, and (c) $0-t_{\text{d2}}$, (d)
$t_{\text{d2}}-t_{\text{dmax}}$ for Voigt geometry.}
\end{figure}

One can see that even a very weak magnetic field of 10~mT influences
the exciton recombination dynamics in both field orientations. The
three specific stages are separated in Fig.~\ref{fig2} by vertical
dashed lines for both Faraday and Voigt geometries. At the first
stage before $t_{\text{d}1}=1~\mu$s, the PL does not depend on magnetic
field. In order to have a closer look at the following dynamic
stages we present them separately in Fig.~\ref{fig3}. In both cases
of Faraday and Voigt geometries, at the second stage for
$t_{\text{d}}$ between $t_{\text{d1}}$ and $t_{\text{d2}}=20~\mu$s
the magnetic field boosts the PL [Figs.~\ref{fig3}(a)
and~\ref{fig3}(c)], while at the third stage, $t_{\text{d}} >
t_{\text{d2}}$, the PL intensity decreases with field application
[Figs.~\ref{fig3}(b) and~\ref{fig3}(d)].

The effect of the magnetic field on the PL decay can be estimated
quantitatively, because it is characterized by the difference of PL intensities, $I_{PL}$, integrated between $t_{\text{d1}}$ and $t_{\text{d2}}$ for the second stage and between $t_{\text{d2}}$ and the maximum delay $t_{\text{dmax}}$ for the third stage, with applied field and without the field. To compare the cases of longitudinal and transverse
magnetic field, these differences have to be normalized by the
integrated intensity of the PL without field as follows:
\begin{eqnarray}
  \label{Eq1}
  S2=\frac{\int_{t_{\text{d1}}}^{t_{\text{d2}}} I_{PL}(t,B)dt - \int_{t_{\text{d1}}}^{t_{\text{d2}}} I_{PL}(t,B=0)dt}{\int_{t_{\text{d1}}}^{t_{\text{d2}}}
  I_{PL}(t,B=0)dt},\\
  S3=\frac{\int_{t_{\text{d2}}}^{t_{\text{dmax}}} I_{PL}(t,B)dt - \int_{t_{\text{d2}}}^{t_{\text{dmax}}} I_{PL}(t,B=0)dt}{\int_{t_{\text{d2}}}^{t_{\text{dmax}}} I_{PL}(t,B=0)dt}.\nonumber
\end{eqnarray}
The calculations show that S2 equals $85$ and $325$, and S3 equals
$-77$ and $-170$ for Faraday and Voigt geometry, respectively.
Note that the magnetic field induces mostly a change in dymanics, while the
decrease of the total PL intensity integrated from $t=0$ to
$t_{\text{dmax}}$ in magnetic field is small, amounting to 1.8$\%$ in
Faraday and 5.0$\%$ in Voigt geometries.

\subsection{Magnetic-field-induced circular polarization of photoluminescence}

\begin{figure}[t]
\centering
\includegraphics[width=1.0\linewidth]{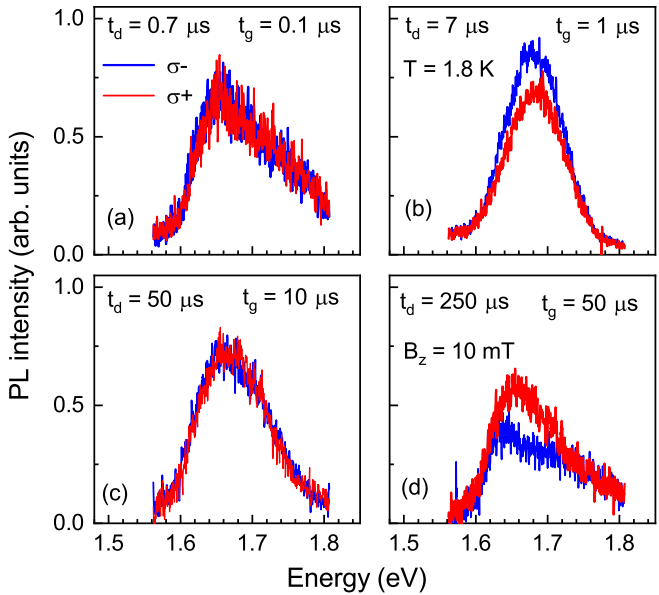}
\caption{PL spectra measured for opposite circular polarizations in
longitudinal magnetic field $B_z=10$~mT, for the experimental settings: (a)
$t_{\rm d}=0.7$~$\mu$s and $t_{\rm g}=0.1$~$\mu$s; (b) $t_{\rm
d}=7$~$\mu$s and $t_{\rm g}=1$~$\mu$s; (c) $t_{\rm d}=50$~$\mu$s and
$t_{\rm g}=10$~$\mu$s; and (d) $t_{\rm d}=250$~$\mu$s and $t_{\rm
g}=50$~$\mu$s.} \label{fig4}
\end{figure}

A very weak (in the few millitesla range) longitudinal magnetic field
leads to the appearance of dynamic electron spin
polarization~\cite{Smirnov125}, which can be evidenced in the circular
polarization of the photoluminescence. At fixed magnetic field, the
sign of the polarization and the polarization degree depend on the delay
after the excitation pulse. This is shown in Fig.~\ref{fig4}, which
demonstrates PL spectra measured at $B_z=10$~mT in $\sigma^+$ and
$\sigma^-$ polarization for different sets of $t_{\text{d}}$ and $t_{\text{g}}$. One can see that the PL is unpolarized at $t_{\text{d}} =$ 0.7~$\mu$s, gets negatively polarized at $t_{\text{d}}=7~\mu$s, again loses polarization at $t_{\text{d}}=50~\mu$s, but becomes
positively polarized at $t_{\text{d}}=250~\mu$s.

\begin{figure}[t]
\centering
\includegraphics[width=0.8\linewidth]{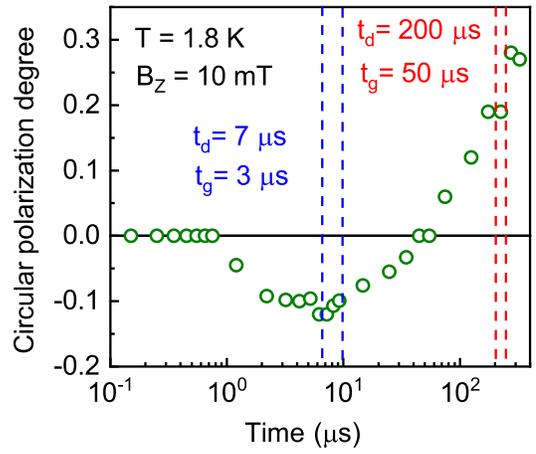}
\caption{Dynamics  of  PL  circular  polarization degree measured
at the energy of $1.67$~eV for the magnetic field $B_z=10$~mT in Faraday
geometry. Vertical lines show the time-integration windows for
Fig.~\ref{fig6}.} \label{fig5}
\end{figure}

The dynamics  of the PL circular polarization degree detected at the
energy of $1.67$~eV for $B_z=10$~mT is shown in Fig.~\ref{fig5}. The negative polarization appears at a delay of about 1~$\mu$s after the pump pulse, the polarization degree decreases
and reaches $-0.12$ at the delay of 8~$\mu$s,  then increases to zero
for a delay of 50~$\mu$s. At large delays, the polarization changes
sign (becomes positive), the polarization degree increases
monotonically and finally reaches 0.28 at the delay of 300~$\mu$s.

\begin{figure}[t]
\centering
\includegraphics[width=0.8\linewidth]{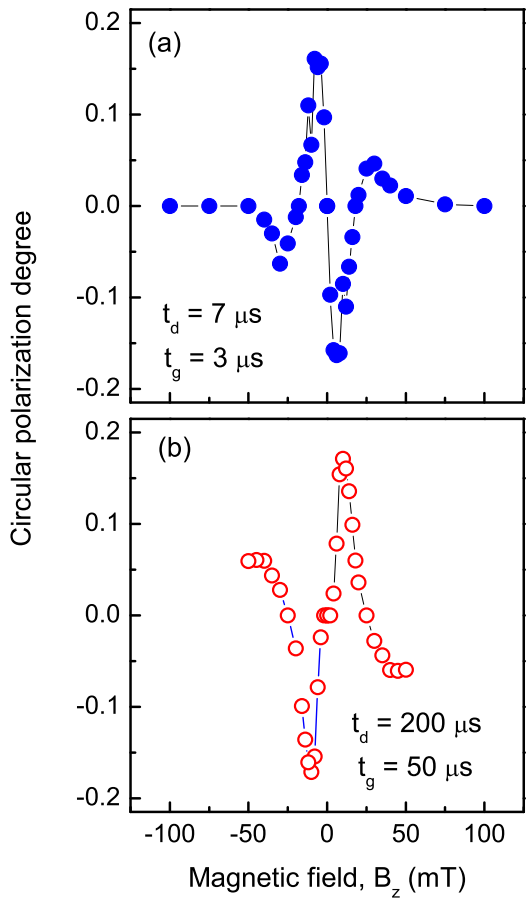}
\caption{Circular polarization degree of the PL at the energy of
$1.66$~eV integrated in the time windows shown in Fig.~\ref{fig5} in
Faraday geometry as a function of magnetic field: (a) $t_{\rm
d}=7$~$\mu$s and $t_{\rm g}=3$~$\mu$s, (b) $t_{\rm d}=200$~$\mu$s
and $t_{\rm g}=50$~$\mu$s.} \label{fig6}
\end{figure}

The magnetic field dependencies of the polarization degree of the PL
integrated for the two time windows shown by the vertical dashed lines in
Fig.~\ref{fig5}, which correspond to negative and positive
PL polarizations are shown in Figs.~\ref{fig6}(a)
and~\ref{fig6}(b), respectively. First of all, it should be noted
that both dependencies are: (i) odd as function of the magnetic field and (ii) strongly
non-monotonic, almost quasi-oscillatory.

For the parameters $t_{\rm d}=7$~$\mu$s and $t_{\rm g}=3$~$\mu$s
corresponding to the negative circular polarization range of the PL
[Fig.~\ref{fig6}(a)] the polarization degree decreases to $-0.16$ in
$B_z=6$~mT, and then increases to zero, changes sign, reaches
0.06 at 30~mT field, and then drops to zero at
fields larger than 50~mT. For the delay range corresponding to the
positive circular polarization of PL at $10$~mT ($t_{\rm
d}=200$~$\mu$s and $t_{\rm g}=50$~$\mu$s) [Fig.~\ref{fig6}(b)], the
$P_c(B)$ shows a qualitatively similar behavior.

\begin{figure}[]
\centering
\includegraphics[width=0.8\linewidth]{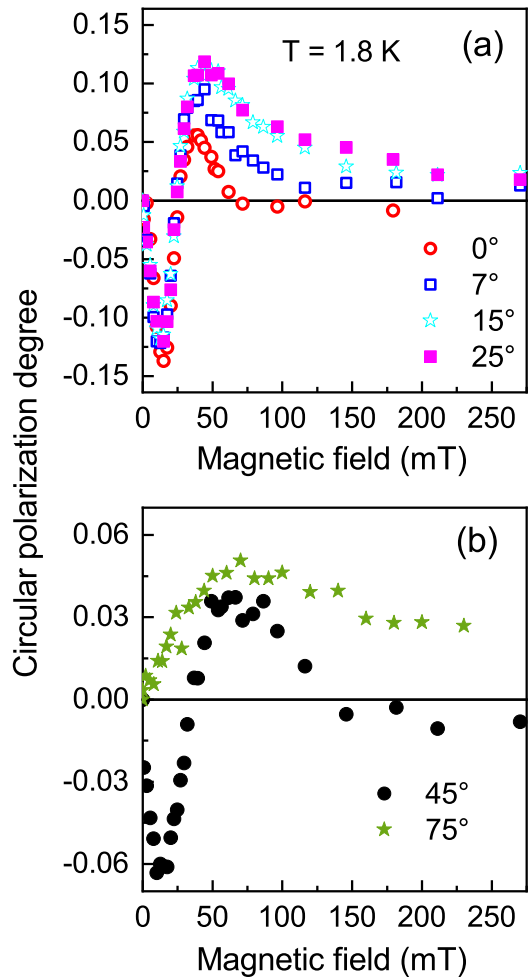}
\caption{Circular polarization degree of PL at the energy of
$1.70$~eV integrated in the time window with $t_{\rm d}=5$~$\mu$s
and $t_{\rm g}=5$~$\mu$s as function of magnetic field for
different angles between the magnetic field direction and the QD
growth axis, $\theta$: (a) 0$^{\circ}$ - Faraday geometry (open red
 circles), 7$^{\circ}$ (open blue squares), 15$^{\circ}$ (open cyan stars) and 25$^{\circ}$ (filled magenta squares); (b)
 45$^{\circ}$ (black circles), and 75$^{\circ}$ (green stars).} \label{fig7}
\end{figure}

A change in the angle $\theta$ between the magnetic field direction
and the QD growth axis modifies the dependence of the PL circular
polarization degree on the magnetic field strength. The $P_c(B)$
measured for the settings $t_{\rm d}=5$~$\mu$s and $t_{\rm
g}=5$~$\mu$s, corresponding to the negative circular polarization range of
the PL, are shown in Figs.~\ref{fig7}(a) and~\ref{fig7}(b) for different
angles $\theta$ in the range $0^{\circ} \leq \theta \leq
75^{\circ}$. One can see that in the angle range $0^{\circ} \leq
\theta \leq 25^{\circ}$ the negatively polarized part of the
$P_c(B)$ curve practically does not change, while for the positively
polarized PL the polarization degree increases and the maximum of
$P_c(B)$ shifts towards stronger magnetic fields. A further
increase of the $\theta$ angle leads to a decrease of the
polarization degree of both the negatively and positively polarized PL.
However, the decrease of the polarization degree of the negatively
polarized PL is faster than that of the positively polarized one.
In fact, the negatively polarized part of $P_c(B)$ completely disappears
for $\theta$ above $60^{\circ}$, while we
observe positively polarized PL up to angles of $75^{\circ}$.

Let us summarize the most important experimental findings:

(i) The application of a weak magnetic field (in both Faraday and
Voigt geometry), modifies the dynamics of the unpolarized PL resulting
in accelerated or decelerated decay in different time ranges,
but does not change the integral of the PL intensity. The effect is similar for
both field orientations, but more pronounced in the Voigt geometry.

(ii) A weak longitudinal magnetic field (Faraday geometry) induces
a circular polarization of the photoluminescence.

(iii) The dynamics of this polarization degree is strongly
non-monotonic in time. After the excitation pulse it first equals to zero, then
the polarization degree becomes negative, drops subsequently again
to zero, changes it sign (from negative to positive) and finally
strongly increases.

(iv) For any delay time after the excitation pulse, either with positive or negative PL polarization, the polarization degree is an odd function of the magnetic field and
varies strongly non-monotonic with magnetic field strength. The polarization degree
increases in very small fields up to a maximum value, then with
rising field decreases to zero, changes its sign and
increases again. A further increase in field strength again leads to
a drop of the polarization degree down to zero.

(v) A change of the angle $\theta$ between the magnetic field
direction and the QD growth axis modifies the dependence of the PL
circular polarization degree on the magnetic field strength. The
positively polarized PL is maintained up to large angles $\theta$,
while the negatively polarized PL observed for $\theta=0^{\circ}$ disappears in
weak magnetic fields.

\section{Theory}
\label{sec:theory}

In this section we develop a microscopic theory of the dynamic electron
spin polarization in quantum dots and its detection through
polarization resolved exciton photoluminescence. We present an
extension of our model from Ref.~\onlinecite{Smirnov125},
considering first a set of identical QDs. The luminescence of
different QDs is considered to be independent. In the next
Sec.~\ref{sec:compare} a set of different QDs will be considered to
describe the experimental results.

\subsection{Model}
\label{sec:model}

An exciton consists of a heavy hole with the spin projections
$J_z=\pm3/2$ along the structure growth axis $z$ and an electron
with the spin projections $S_z=\pm1/2$. We account for the external
magnetic field, the hyperfine interaction between the electron spin
and the unpolarized spins of the nuclei in the QD, and the short
range exchange interaction between the electron and hole spins.

\begin{figure}[t]
\centering
\includegraphics[width=0.9\linewidth]{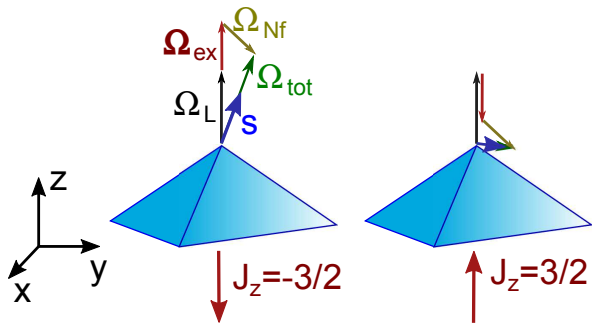}
\caption{Electron spin precession frequency $\bm\Omega_{\rm tot}$
[Eq.~\eqref{eq:Omega_tot}] and the contributions to it for the two
heavy hole spin orientations. The average electron spin $\bm S$ is
parallel to $\bm\Omega_{\rm tot}$ due to fast precession.}
\label{fig8}
\end{figure}

The system Hamiltonian can be written as
\begin{equation}
  \mathcal H=\hbar\bm\Omega_{\rm tot}\bm S,
\end{equation}
where
\begin{equation}
  \label{eq:Omega_tot}
  \bm\Omega_{\rm tot}=\bm\Omega_L+\bm\Omega_{\rm Nf} +  \bm\Omega_{\rm ex}
\end{equation}
is the total electron spin precession frequency. Here, $\bm
\Omega_L$ is the electron Larmor spin precession frequency in the
external magnetic field $\bm B=\hbar\bm\Omega_L/(g_e\mu_B)$ with
$g_e$ being the electron $g$ factor and $\mu_B$ being the Bohr
magneton, $\bm\Omega_{\rm Nf}$ is the precession frequency in the
Overhauser field of the randomly oriented nuclear spins in the QD,
$\bm\Omega_{\rm ex} = -\frac{2}{3}\delta_0J_z \bm{e}_z/\hbar$ is the
precession frequency in the exchange field of the heavy hole, with
$\delta_0$ being the short range exchange interaction
constant~\cite{Ivch_book,Astakhov} and $\bm{e}_z$ being the unit
vector along the $z$ axis. Noteworthy, due to this term the total
electron spin precession frequency depends on the hole spin $J_z$,
as it is shown in Fig.~\ref{fig8}.

For simplicity we neglect a number of contributions in the
Hamiltonian. First, we neglect the long range electron hole exchange
interaction, which is suppressed for the indirect
excitons~\cite{Bir,Goupalov,Kuznetsova} as well as possible
noncollinear terms in the short range exchange interaction for electrons in the $X$ valley. We also neglect the hole Zeeman splitting, because the transverse components
of the tensor of the heavy hole $g$ factors are very
small~\cite{Debus,Golub}, while the longitudinal component does not
play a role in the phenomena addressed here~\cite{Smirnov125}.
For the same reasons we neglect the
hyperfine interaction for the hole, which is additionally suppressed by
the p-type orbitals of the Bloch wave functions~\cite{Glazov,Chekhovich}. For
the electrons we neglect the possible valley degeneracy and assume
that the electron spin dynamics take place only within one of the
valleys, while in general the hyperfine interaction can scatter
electrons between the valleys~\cite{Avdeev}, which can lead to
spin relaxation. For simplicity we neglect the nuclear spin
polarization~\cite{OO_book,Korenev} and nuclear spin dynamics, which
in principle can take place at submillisecond time
scale~\cite{Petrov,Inertia,CSM}. Further, we neglect the anisotropy
of the hyperfine interaction~\cite{Shchepetilnikov,Kuznetsova}, and
assume that the distribution function of the Overhauser field has
the form~\cite{Merkulov,PRC_general}:
\begin{equation}
  \mathcal F({\bm B}_{\rm Nf}) = \frac{1}{(\sqrt{\pi}\Delta_B)^{3}}\exp\left(-\frac{{B}_{\rm Nf}^2}{\Delta_B^2}\right),
\end{equation}
where the parameter $\Delta_B$ characterizes the dispersion of the
Overhauser field. Thus, the exciton spin dynamics due to the
hyperfine interaction is reduced to precession with a static
frequency $\bm{\Omega}_{\rm Nf} = \mu_B g_e {\bm B}_{\rm Nf} /\hbar
$, and the spin dynamics should be averaged over the distribution
$\mathcal F({\bm B}_{\rm Nf})$.

\begin{figure}[t]
\centering
\includegraphics[width=0.9\linewidth]{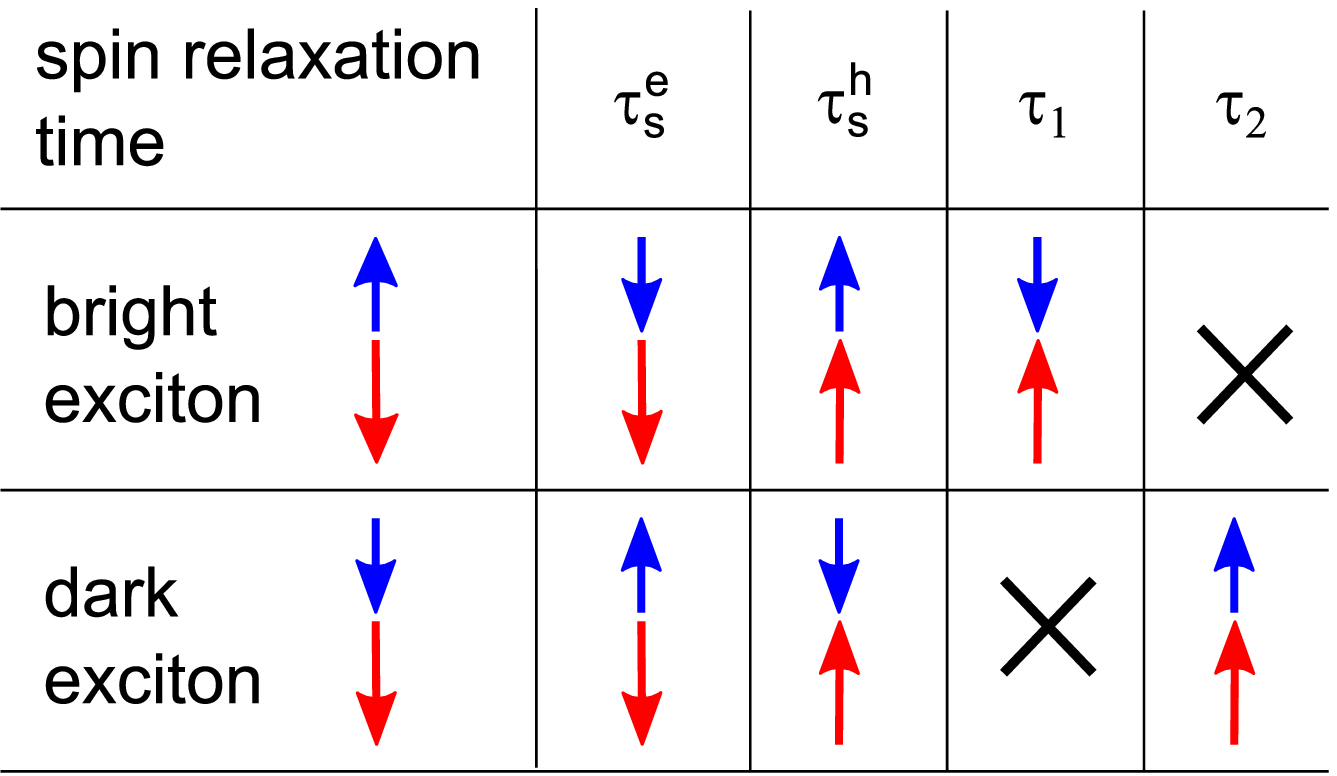}
\caption{The action of the different spin relaxation mechanisms
described by the times $\tau_s^e$, $\tau_s^h$, $\tau_1$, and
$\tau_2$ on the electron (blue arrow) and hole (red arrow) spin
projections onto the $z$ axis inside the bright and dark excitons.
The initial states and the results of the action are shown in the
first and the corresponding columns, respectively. The spin
relaxations described by the times $\tau_1$ and $\tau_2$ do not
affect the dark and bright states, respectively.} \label{fig9}
\end{figure}

Since $J_z$ is a good quantum number, it is useful to introduce
$N^{(+)}$ and $N^{(-)}$ as the numbers of excitons with hole spin
projections $+3/2$ and $-3/2$, respectively~\cite{relations}. We also introduce the
averaged spin polarizations ${\bm S}^{(\pm)}$ of the electrons in
the excitons with $J_z = \pm 3/2$. Then the exciton
dynamics including incoherent processes is
described by the following equations:
\begin{widetext}
  \begin{subequations}
    \label{eq:all}
    \begin{multline}\label{eqN}
      \frac{d N^{(\pm)}}{dt} = -\frac{1}{\tau_R} \left(\frac{N^{(\pm)
          }}{2} \mp S_z^{(\pm) } \right) -\frac{1}{\tau_{NR} }
      \left(\frac{N^{(\pm) }}{2} \pm S_z^{(\pm) } \right) +
      \frac{N^{(\mp)} - N^{(\pm)}}{2\tau_s^h}
      + \left( \frac{1}{\tau_1} + \frac{1}{\tau_2} \right) \frac{N^{(\mp)} - N^{(\pm)}}{4} \\
      \pm \left( \frac{1}{\tau_1} - \frac{1}{\tau_2} \right)\frac{S_z^{(+)} + S_z^{(-)}}{2},
    \end{multline}
    \begin{multline}\label{eqSz}
      \frac{d S_{z}^{(\pm)}}{dt} = \pm\left( \frac{1}{\tau_R} - \frac{1}{\tau_{NR}} \right) \frac{N^{(\pm)}}{4} - \left( \frac{1}{\tau_s^e} + \frac{1}{2\tau_{R}} + \frac{1}{2\tau_{NR}}  \right)S_{z}^{(\pm)} + \frac{S_{z}^{(\mp)} - S_{z}^{(\pm)}}{2\tau_s^h} + [\bm{\Omega}_{\rm tot}^{(\pm)} \times {\bm S}^{(\pm)}]_{z}\\
      -\left( \frac{1}{\tau_1} + \frac{1}{\tau_2} \right)
      \frac{S_z^{(+)} + S_z^{(-)}}{4} + \left( \frac{1}{\tau_1} -
        \frac{1}{\tau_2} \right) \frac{N^{(+)} - N^{(-)}}{8},
    \end{multline}
    \begin{equation}\label{eqSxy}
      \frac{d S_{x,y}^{(\pm)}}{dt} = - \left( \frac{1}{\tau_s^e} + \frac{1}{2\tau_{R}} + \frac{1}{2\tau_{NR}}  + \frac{1}{2\tau_1} + \frac{1}{2\tau_2} \right)S_{x,y}^{(\pm)}
      + \frac{S_{x,y}^{(\mp)} - S_{x,y}^{(\pm)}}{2\tau_s^h}  + [\bm{\Omega}_{\rm tot}^{(\pm)} \times {\bm S}^{(\pm) }]_{x,y}.
    \end{equation}
  \end{subequations}
\end{widetext}
Here ${\bm \Omega}_{\rm tot}^{(\pm)}=\bm\Omega_L+\bm\Omega_{\rm
Nf}\mp \delta_0\bm e_z/\hbar$ is the vector frequency of the
electron spin precession corresponding to the hole spin projections
$J_z = \pm 3/2$, $\tau_{s}^{e}$ and $\tau_{s}^{h}$ are the electron
and hole spin relaxation times, respectively, $\tau_{1}$ and
$\tau_{2}$ are the bright and dark excitons spin flip times, and
$\tau_{R}$ and $\tau_{NR}$ are the lifetimes of the bright and dark
excitons related to radiative and nonratiative recombination. The
action of the spin relaxation terms is illustrated in
Fig.~\ref{fig9}.

Pulsed unpolarized optical excitation is described by the initial
conditions $N^{(\pm)}=N_0/2$ with $N_0$ being the number of generated
excitons and $\bm S^{(\pm)}=0$. The PL intensities in
$\sigma^\pm$ polarization are proportional to
\begin{equation}
R_\pm = \frac{N^{(\pm)}/2\mp S_z^{(\pm)}}{\tau_R},
\end{equation}
respectively. The degree of the circular polarization is given by
\begin{equation}
P = \frac{\braket{R_+ - R_-}}{\braket{R_+ + R_-}},
\end{equation}
where the angular brackets denote the averaging at the given time
over the random nuclear fields.

The electron spin precession is typically much faster than all the
the incoherent processes. So we use the adiabatic approximation and
replace the electron spins $\bm S^{(\pm)}$ with their values
averaged over the precession period, which are parallel to
$\bm\Omega_{\rm tot}^{(\pm)}$. Then, it is useful to introduce
\begin{equation}
{\cal B}^{(\pm)} = \frac{1}{2} N^{(\pm)} \mp {\bm S}^{(\pm)} {\bm e}_\Omega^{(\pm)},
\end{equation}
\begin{equation}
{\cal D}^{(\pm)} = \frac{1}{2} N^{(\pm)} \pm {\bm S}^{(\pm)} {\bm e}_\Omega^{(\pm)},
\end{equation}
with ${\bm e}_\Omega^{(\pm)} = {\rm sign}\left[(\Omega_{\rm
tot}^{(\pm)})_z\right]\cdot {\bm \Omega}_{\rm tot}^{(\pm)}/|{\bm
\Omega}_{\rm tot}^{(\pm)}|$ being the unit
vector along $\bm\Omega_{\rm tot}^{(\pm)}$ with positive $z$
component. So ${\cal B}^{(\pm)}$ and ${\cal D}^{(\pm)}$ have the
meaning of the numbers of mainly bright and mainly dark excitons with
the electron spin being parallel to its precession frequency. If ${\bm
\Omega}_{\rm tot}$ is parallel to the $z$ axis, they are equal to the
number of the corresponding bright and dark excitons. These
excitons are called quasi bright and quasi dark, respectively, in the
following.

For infinite $\tau_{1}$ and $\tau_{2}$, we obtain the
following kinetic equations for the numbers of quasi bright and quasi dark
excitons:
\begin{widetext}
\begin{equation} \label{kB_basic}
\frac{d {\cal B}^{(\pm)}}{dt} =
 - \left( \frac{1+\cos\alpha_\pm}{\tau_R} + \frac{1-\cos\alpha_\pm}{\tau_{NR}} + \frac{1}{\tau_s^e} + \frac{1}{\tau_s^h} \right) \frac{{\cal B}^{(\pm)}}{2}
  + \frac{{\cal D}^{(\pm)}}{2\tau_s^e} + \frac{1-\cos\beta}{4\tau_s^h} {\cal B}^{(\mp)}
  + \frac{1+\cos\beta}{4\tau_s^h} {\cal D}^{(\mp)},
\end{equation}
\begin{equation}\label{kD_basic}
\frac{d {\cal D}^{(\pm)}}{dt} = - \left( \frac{1-\cos\alpha_\pm}{\tau_R} + \frac{1+\cos\alpha_\pm}{\tau_{NR}} + \frac{1}{\tau_s^e} + \frac{1}{\tau_s^h} \right)\frac{{\cal D}^{(\pm)}}{2}
  + \frac{{\cal B}^{(\pm)}}{2\tau_s^e} + \frac{1+\cos\beta}{4\tau_s^h} {\cal B}^{(\mp)}
  + \frac{1-\cos\beta}{4\tau_s^h}{\cal D}^{(\mp)}.
\end{equation}
%\end{widetext}
Here $\alpha_\pm$ is the angle between the axes of ${\bm
\Omega}_{\rm tot}^{(\pm)}$ and the $z$ axis, and $\beta$
is the angle between the axes of ${\bm \Omega}_{\rm tot}^{(\pm)}$.
Importantly, these are the angles between the axes and not between
the directions, so $0\le\alpha_\pm$ and $\beta\le\pi/2$. We
note that after a hole spin flip occuring on the typical time scale of $\tau_s^h$,
the quasi bright exciton with $J_z = +3/2$ becomes an exciton with
hole spin $-3/2$ and since the effective fields ${\bm
\Omega}_{\rm tot}^{(\pm)}$ are different, the exciton remains quasi
bright with the probability $(1-\cos\beta)/2$ and
becomes quasi dark with the probability
$(1+\cos\beta)/2$. The terms corresponding to the other
processes can be described in the same way.

The terms related to $\tau_1$ and $\tau_2$ give the following
contributions to ${d {\cal B}^{(\pm)}}/{dt}$ and ${d {\cal
D}^{(\pm)}}/{dt}$, respectively:
%\begin{widetext}
\begin{subequations}
  \label{eq:kBDtau}
  \begin{multline}\label{kBtau}
    \left(\frac{d {\cal B}^{(\pm)}}{dt}\right)' = - \left( \frac{1}{\tau_1}  - \frac{(1-\cos\alpha_\pm)^2}{4\tau_1} + \frac{1}{\tau_2}  -  \frac{(1+\cos\alpha_\pm)^2}{4\tau_2} \right)\frac{{\cal B}^{(\pm)}}{2} + \left( \frac{1}{\tau_1} + \frac{1}{\tau_2} \right) \frac{\sin^2\alpha_\pm}{8} {\cal D}^{(\pm)} \\
+ \frac{(1+\cos\alpha_\pm)(1+\cos\alpha_\mp)}{8\tau_1}{\cal B}^{(\mp)}  +\frac{(1-\cos\alpha_\pm)(1-\cos\alpha_\mp)}{8\tau_2}{\cal B}^{(\mp)} \\
+ \frac{(1+\cos\alpha_\pm)(1-\cos\alpha_\mp)}{8\tau_1}{\cal D}^{(\mp)}  +\frac{(1-\cos\alpha_\pm)(1+\cos\alpha_\mp)}{8\tau_2}{\cal D}^{(\mp)},
  \end{multline}
  \begin{multline}\label{kDtau}
    \left(\frac{d {\cal D}^{(\pm)}}{dt}\right)' = - \left( \frac{1}{\tau_1}  - \frac{(1+\cos\alpha_\pm)^2}{4\tau_1} + \frac{1}{\tau_2}  -  \frac{(1-\cos\alpha_\pm)^2}{4\tau_2} \right)\frac{{\cal D}^{(\pm)}}{2} + \left( \frac{1}{\tau_1} + \frac{1}{\tau_2} \right) \frac{\sin^2\alpha_\pm}{8} {\cal B}^{(\pm)}  \\
+ \frac{(1-\cos\alpha_\pm)(1+\cos\alpha_\mp)}{8\tau_1}{\cal B}^{(\mp)}  +\frac{(1+\cos\alpha_\pm)(1-\cos\alpha_\mp)}{8\tau_2}{\cal B}^{(\mp)}\\
+ \frac{(1-\cos\alpha_\pm )(1-\cos\alpha_\mp)}{8\tau_1}{\cal D}^{(\mp)}   +\frac{(1+\cos\alpha_\pm)(1+\cos\alpha_\mp)}{8\tau_2}{\cal D}^{(\mp)}.
  \end{multline}
\end{subequations}
%Eqs.~(\ref{kDtau},\ref{kBtau}) contain only the terms related to exciton spin relaxation that should be added to Eqs.~(\ref{kD_basic},\ref{kB_basic}).
\end{widetext}
%\commentDima{Andrei, please, check my corrections in these equations!} \commentAndrei{Dima, you cannot combine terms with ${\cal B}^{\mp}$ and ${\cal D}^{\mp}$ in Eqs.~(12). Check the signs carefully.}
To interpret these  contributions, let us consider, for example, the
terms in Eqs.~\eqref{eq:kBDtau} that are related to the spin flip of
the dark exciton with $J_z = +3/2$ described by the time $\tau_2$.
One can say that with the rate $1/(2\tau_2)$ it becomes either dark
with the probability $(1+\cos\alpha_+)/2$ or bright with the
probability $(1-\cos\alpha_+)/2$. If it became dark, it has changed
the direction of both the electron and hole spins. Then due to the
fast electron spin precession it contributes to ${\cal D}^{(-)}$
with the probability $(1+\cos\alpha_-)/2$ or to ${\cal B}^{(-)}$
with the probability $(1-\cos\alpha_-)/2$. At the same time, the
bright exciton does not change its spin because $\tau_2$ is relevant
for the dark excitons only, so it contributes to ${\cal D}^{(+)}$
with the probability $(1-\cos\alpha_+)/2$ and to ${\cal B}^{(+)}$
with the probability $(1+\cos\alpha_+)/2$. The rest of the terms can
be interpreted in the same way.

The circularly polarized PL components in this limit are
described by
\begin{equation}
R_{\pm} = \frac{1+\cos\alpha_\pm}{2\tau_R} {\cal B}^{(\pm)}+ \frac{1-\cos\alpha_\pm}{2\tau_R} {\cal D}^{(\pm)}.
\end{equation}

It is instructive to consider the absence of hole spin relaxation as
well as bright and dark exciton spin flips. In this case, the hole
spin is conserved, so Eqs.~\eqref{kD_basic} and~\eqref{kB_basic} are
reduced to the two independent sets of equations:
\begin{equation}\label{init}
\frac{d}{dt} \left( \begin{array}{c} {\cal D}^{(\pm)} \\ {\cal B}^{(\pm)}\end{array} \right)
= \left(-\frac{1}{2\tau_0} -\frac{1}{2\tau_s^e} + M^{\pm 3/2}\right)
 \left( \begin{array}{c} {\cal D}^{(\pm)} \\ {\cal B}^{(\pm)}\end{array} \right),
\end{equation}
with the matrices
\begin{equation}
M^{\pm 3/2} = \left(
\begin{array}{cc}
\frac{1}{2\tau_\alpha} \cos\alpha_{\pm}  & \frac{1}{2\tau_s} \\
 \frac{1}{2\tau_s} & -\frac{1}{2\tau_\alpha} \cos\alpha_{\pm}
\end{array}
\right).
\end{equation}
Here $\tau_0 = \tau_R\tau_{NR}/(\tau_R + \tau_{NR})$ and $\tau_\alpha = \tau_R\tau_{NR}/(\tau_{NR} - \tau_{R})$. These matrices have the eigenvalues $\gamma_{\pm}$ and $-\gamma_{\pm}$, respectively, where $\gamma_{\pm} = \sqrt{\tau_s^2 \cos^2\alpha_{\pm} + \tau_\alpha^2}/2\tau_s \tau_\alpha$. Therefore, the PL dynamics is biexponential:
\begin{multline} \label{2LS-int}
R_{\pm}(t) = \exp\left(-\frac{t}{2\tau_s} - \frac{t}{2\tau_\alpha}\right)\times \\
\left(R_\pm^{(1)} e^{-\gamma_\pm t} + R_\pm^{(2)} e^{\gamma_\pm t}\right),
\end{multline}
where
\begin{equation}
R_\pm^{(1)} = \frac{1}{2\tau_R} \left[
\frac{(\xi_\pm + \tau_\alpha)^2 }{\tau_\alpha^2 + \xi_\pm^2} +
\frac{\tau_\alpha^2 - \xi_\pm^2 }{\tau_\alpha^2 + \xi_\pm^2} \cos\alpha_\pm
\right],
\end{equation}
\begin{equation}
R_\pm^{(2)} = \frac{1}{2\tau_R} \left[
\frac{(\zeta_\pm + \tau_\alpha)^2 }{\tau_\alpha^2 + \zeta_\pm^2} +
\frac{\tau_\alpha^2 - \zeta_\pm^2 }{\tau_\alpha^2 + \zeta_\pm^2} \cos\alpha_\pm
\right]
\end{equation}
with $\xi_\pm = \tau_s(\cos\alpha_{\pm} - 2\tau_\alpha \gamma_\pm)$ and $\zeta_\pm = \tau_s(\cos\alpha_{\pm} - 2\tau_\alpha \gamma_\pm)$.

Generally, Eq.~\eqref{2LS-int} has to be averaged over the
distribution of the nuclear fields. If, however, the external
magnetic field $\Omega_L$ is larger then the exchange field
$\Omega_{ex}$ and is tilted from the $z$ axis by the angle $\theta
\ne 0$, its transverse component takes a role similar to the
nuclear field fluctuations. If the latter are smaller, then one can
take
\begin{equation}\label{highH}
  \cos\alpha_{\pm} = \cos\theta \pm \frac{|\Omega_{ex}|}{\Omega_L}(1-\cos\theta).
\end{equation}
This gives us an analytical expression for the PL polarization.

\begin{figure}[t]
\centering
\includegraphics[width=0.95\linewidth]{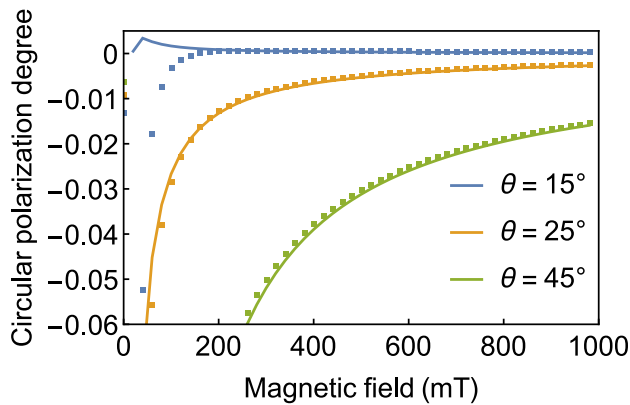}
\caption{PL polarization degree calculated analytically after
Eqs.~\eqref{2LS-int} and~\eqref{highH} (the solid curves) and
numerically after Eqs.~\eqref{eq:all} (the symbols) for the
parameters $\Delta_B = 10$~mT, $B_{ex} = \hbar |\Omega_{ex}|
/(g\mu_B)= 5$~mT, $\tau_R = 1~\mu$s, $\tau_{NR} = 12~\mu$s,
$\tau_s^e = 15~\mu$s, $\tau_s^h = \tau_1=\tau_2 = 10^3~\mu$s, the
delay time $t_{\rm d} = 20~\mu$s. The numerical averaging was
performed over 3000 hyperfine field realizations.} \label{fig10}
\end{figure}

In Fig.~\ref{fig10} we demonstrate that  the analytical and
numerical results agree well for large magnetic fields. An increase
of the magnetic field leads to the suppression of the dynamic
polarization. However, the larger the tilt angle of the field, the
stronger the polarization.

\subsection{Numerical results}
\label{sec:num}

In this section we use  numeric calculations to describe the
dependence of the PL intensity and polarization on time as well as
magnetic field strength and orientation. We average the numeric
solution of Eqs.~\eqref{eq:all} over $10^4$ hyperfine field
realizations unless stated otherwise.

\begin{figure}[t]
\centering
\includegraphics[width=0.95\linewidth]{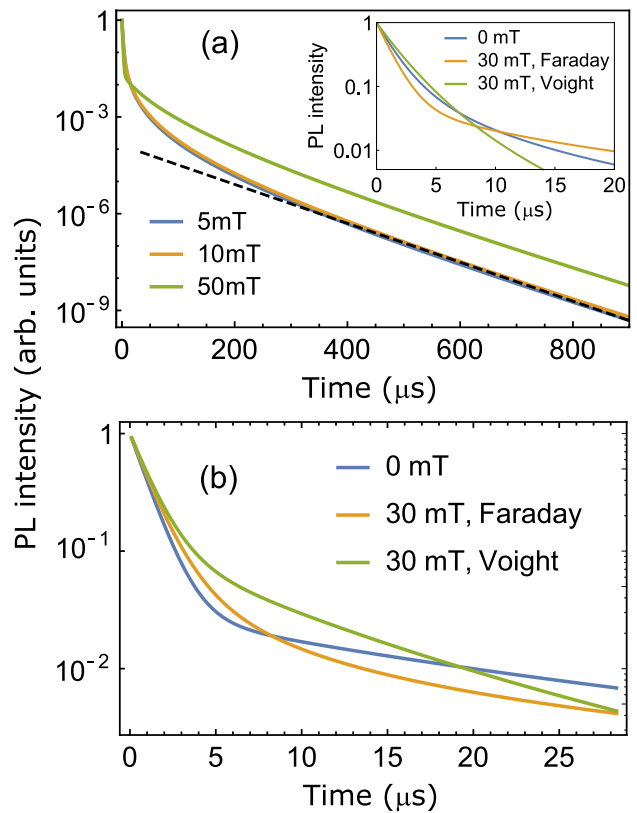}
\caption{(a) Dynamics of the unpolarized PL calculated for different
magnetic fields in Faraday geometry using the parameter $\Delta_B =
B_{ex} = 10 \, {\rm mT}$. The black dashed line is proportional to
$\exp(-t/ 72 \, \mu{\rm s})$. The inset shows the unpolarized PL
dynamics calculated at short time scales for zero field and a field
of $30 \,{\rm mT}$ in Faraday and Voigt geometry. (b) Unpolarized PL
dynamics calculated using the parameters $\Delta_B  = 10 \, {\rm
mT}$, $B_{ex} = 40 \, {\rm mT}$ in Faraday and Voigt geometry at
short time scales. The other parameters are given in text.}
\label{fig11}
\end{figure}

\begin{figure}[t]
\centering
\includegraphics[width=0.95\linewidth]{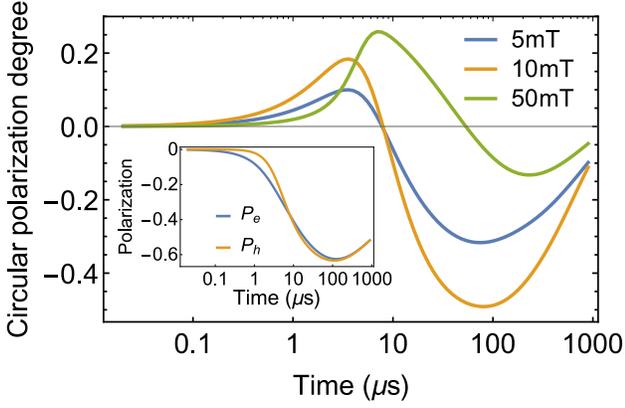}
\caption{The dynamics of the PL polarization degree calculated for
$\Delta_B = B_{ex} = 10 \, {\rm mT}$ in different magnetic fields.
The inset shows the electron and hole spin polarization at $B =
10\,{\rm mT}$.} \label{fig12}
\end{figure}

We start the discussion using the following set of parameters
$\Delta_B = B_{ex} = 10\,{\rm mT}$, $\tau_R = 1\,\mu {\rm s}$,
$\tau_{NR} = 100\, \mu {\rm s}$, and
$\tau_s^e=\tau_s^h=\tau_1=\tau_2 = 1\,{\rm ms}$. In this case, the
spin relaxation does not play a role. At the same time, the
nonradiative recombination is slow, but it can affect the
polarization at corresponding large times. We consider external
magnetic fields up to 100~mT along the $z$ axis. The electron $g$
factor in indirect band gap (In,Al)As QDs was recently measured to
be $g_e = 2$~\cite{Debus,Ivanov97}.

The dynamics of the unpolarized PL intensity (sum of $\sigma^+$ and
$\sigma^-$ polarized PL components), calculated for different
magnetic fields, is shown in Fig.~\ref{fig11}(a). In a given QD it
consists of four exponential contributions [see
Eq.~\eqref{2LS-int}], and after averaging over the hyperfine fields
it becomes even more complex. However, at long delays the PL
dynamics becomes almost mono-exponential. This is shown in
Fig.~\ref{fig11}(a) by the black dashed line that corresponds well
to a phenomenological PL dependence on time $\propto \exp(-t/ 72 \,
\mu{\rm s})$.

The effect of the magnetic field can vary qualitatively, depending
on the relation between the typical random nuclear field $\Delta_B$
and the exchange field $B_{\rm ex}$. In the inset of
Fig.~\ref{fig11}(a) we show that for $\Delta_B\gtrsim B_{\rm ex}$
the magnetic field accelerates the exciton recombination in the
Voigt geometry, but decelerates it in the Faraday geometry. This is
explained by the different mixing between bright and dark excitons
in these two cases, which is much stronger in the Voigt than in the
Faraday geometry. At the same time, in Fig.~\ref{fig11}(b) we
demonstrate that for $\Delta_B\ll B_{ex}$ the magnetic field can
accelerate the PL decay in the Faraday geometry as well. The reason
for this is the reduced splitting between bright and dark excitons
for one of the heavy hole spin states.

The calculated dynamics of PL polarization in Faraday geometry is
shown in Fig.~\ref{fig12} using the same parameters (in the Voigt
geometry polarization is symmetry forbidden). The polarization is
zero at $t=0$, increases up to $t\sim\tau_R= 1\, \mu {\rm s}$, and
then decreases and changes sign at $t\sim\tau_{NR}= 100\, \mu {\rm
s}$.

The dynamic circular polarization of the PL is induced by the
external magnetic field and it is an odd function of $B_z$. The PL
is positively polarized at short delays, and negatively polarized at
long delays because the mixing between the bright and dark exciton
states with $J_z=+3/2$ is stronger than between those with
$J_z=-3/2$.

For the same reason the electron and hole spins in an exciton
become also dynamically polarized. The electron polarization $P_e$
and hole polarization $P_h$ are defined as follows
\begin{subequations}
\begin{equation}
P_e = 2\frac{\braket{S_z^{3/2}+S_z^{-3/2}}}{\braket{N^{3/2}+N^{-3/2}}},
\end{equation}
\begin{equation}
P_h = \frac{\braket{N^{3/2}-N^{-3/2}}}{\braket{N^{3/2}+N^{-3/2}}}.
\end{equation}
\end{subequations}
The polarizations $P_e$ and $P_h$ are shown in the inset of
Fig.~\ref{fig12} and behave similarly to the PL polarization, but
remain always negative. At long time delays only dark excitons are
left, so the polarizations of the electrons, the holes and the PL
coincide.

In the absence of nonradiative recombination all excitons recombine
radiatively. Since we neglect the heavy hole spin flips, the
excitons with $J_z=\pm3/2$ emit $\sigma^\pm$ polarized light,
respectively. For unpolarized excitation their numbers are equal, so
the integral PL is unpolarized.

\begin{figure}[t]
\centering
\includegraphics[width=0.95\linewidth]{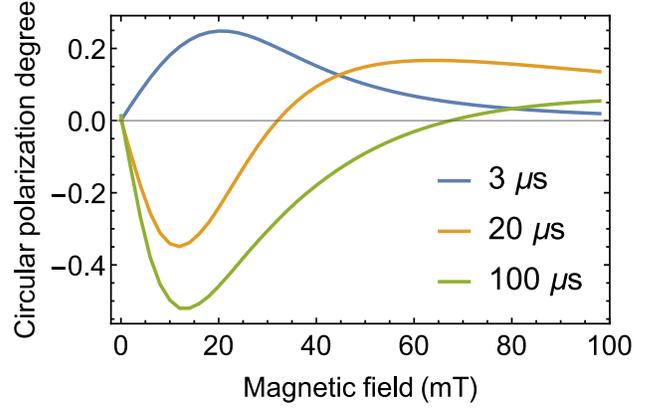}
\caption{Magnetic field dependence of the PL polarization degree
calculated for the different delay times indicated in the figure.}
\label{fig13}
\end{figure}

The PL polarization degree calculated as a function of magnetic
field for a few fixed time delays is shown in Fig.~\ref{fig13}. One
can see that it can change sign from negative to positive with
increasing magnetic field strength. This can be explained as
follows: The polarization is positive at short time delays and
negative at long ones, as shown in Fig.~\ref{fig12}. However, with
increasing field strength in the range $B\gtrsim\Delta_B$ the
positive part of the polarization moves to longer times. For
example, for $B=10\,{\rm mT}$ the polarization changes sign at
$t\approx 8\,{\rm \mu s}$, while for $B=50\,{\rm mT}$ it changes its
sign at $t \approx 54\,\mu{\rm s}$. As a result, the polarization at
a given time changes its sign as a function of magnetic field. For
$t=20 \,{\rm \mu s}$ it happens for the field $B \approx 32 \,{\rm
mT} $, as shown in Fig.~\ref{fig13}.

\begin{figure}[t]
\centering
\includegraphics[width=0.95\linewidth]{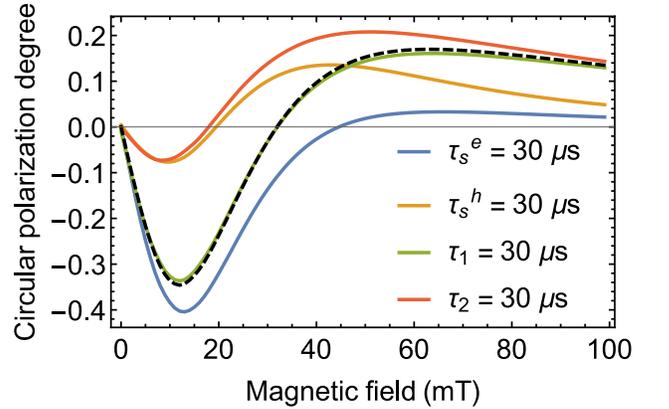}
\caption{ Magnetic field dependencies of the PL polarization degree
calculated for the delay time of $20~\mu$s, accounting for different
spin relaxation times. The black dashed curve reproduces the yellow
curve from Fig.~\ref{fig13}.} \label{fig14}
\end{figure}

The influence of the various spin relaxation  mechanisms is
illustrated in Fig.~\ref{fig14}, where we consider the PL
polarization degree for the time delay of $20~\mu$s. Here, the black
dashed curve reproduces the yellow curve in Fig.~\ref{fig13} and the
other curves show the effects of the different spin relaxation times
being changed from $1$~ms to $30~\mu$s. First, the blue curve
demonstrates that a decrease of the electron spin relaxation time
$\tau_s^e$ leads to suppression of the positive polarization in
large fields. The red and orange curves demonstrate that decreases
of the hole and of the dark exciton spin relaxation times,
$\tau_s^h$ and $\tau_2$, respectively, have a similar effect: they
lead to suppression of the negative polarization in small fields.
Finally, the green curve shows that variation of the bright exciton
spin relaxation time $\tau_1$ have almost no effect, the curve
almost coincides with the black dashed one. Note, that this
behaviour may change for $\tau_1$ shorter than $\tau_R$.

\begin{figure}[t]
\centering
\includegraphics[width=0.95\linewidth]{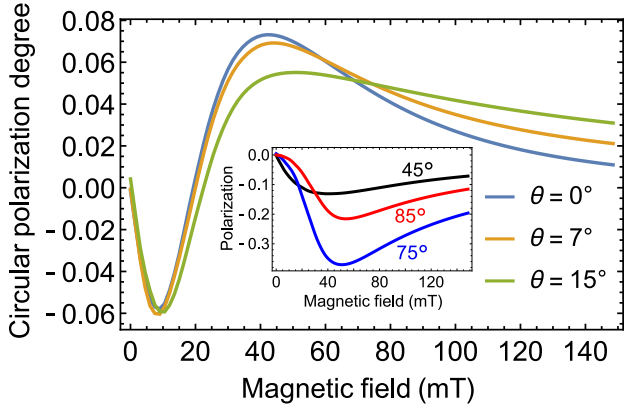}
\caption{Magnetic field dependencies of the PL polarization degree
calculated for $t_{\rm d} = 15\,{\rm {\mu}s}$ and $t_{\rm g} =
30\,{\rm \mu s}$ for different magnetic field tilt angles. The inset
shows the same for large tilt angles indicated at the different
curves.} \label{fig15}
\end{figure}

The effect of a magnetic field tilt angle is illustrated in
Fig.~\ref{fig15}. Here, the exchange field $B_{ex} = 7 \, {\rm mT}$
was used ($\Delta_B = 10\,{\rm mT}$). The electron and dark exciton
spin relaxations were taken to be $\tau_s^e = 40\, {\rm \mu s}$ and
$\tau_2 = 20\,{\rm \mu s}$. The photoluminescence was
time-integrated with the parameters $t_{\rm d} = 15\,{\rm \mu s}$
and $t_{\rm g} = 30 \,{\rm \mu s}$. One can see that at small angles
the low field negative polarization is almost the same as for the
pure Faraday geometry. The tilt angle flattens the positive part of
the polarization. Its decrease with increasing magnetic field gets
less pronounced, in agreement with the results of
Sec.~\ref{sec:model} and Fig.~\ref{fig10}. The inset shows that for
large tilt angles the polarization minimum shifts to larger fields
and becomes deeper. Its position corresponds to $B_z$ of the order
of $\Delta_B$. At the same time, the positive part of the
polarization disappears. This is related to the suppression of
radiative recombination of pseudo-dark excitons at $B_z > \Delta_B$.
Noteworthy, a strongly tilted magnetic field can mix dark and bright
excitons even without hyperfine field and lead to dynamic
polarization.
% Therefore large magnetic field does not suppress the mixing.

Concluding this section, we note that the concept of dynamic
electron spin polarization is similar to that of nuclear
self-polarization~\cite{self_nuclei} and is very general. It can
also take place, for example, when the electrons and holes are
injected into the QDs not optically, but electrically. It is known
that in organic semiconductors the hyperfine and exchange
interaction lead to electron and hole spin correlations even at room
temperature, which can be evidenced in
magnetoresistance~\cite{OMAR0,OMAR-ASh1,OMAR-ASh2} and modification
of the total PL intensity~\cite{OMAR-Kalin,OMAR-Vard} in small
magnetic fields. Since optical orientation is very inefficient in
organic semiconductors, the dynamic spin polarization may be a
useful tool for spin initialization.

\section{Comparison between experiment and theory}
\label{sec:compare}

The theoretical model and numerical calculated results presented in
Sec.~\ref{sec:theory} consider identical QDs, while in the
experiment the size, shape, composition, and
heterointerface smoothness strongly vary in an
ensemble~\cite{Shamirzaev78,Shamirzaev84}. However, the dynamic spin
polarization requires a small exchange interaction, so the
relevant electron states all belong to the $X$ valleys. As a result, the
electron $g$ factor equals to $2$ in all the QDs. Moreover, since
the hyperfine interaction is dominated by the contact interaction
with the As nuclei~\cite{Kuznetsova}, the composition variations in (In,Al)As do not
lead to variations of $\Delta_B$. Further, in experiment we detect
the PL in a rather narrow window $1.66-1.70$~eV, which reduces the
variance of QD sizes and, as a result, of $B_{\rm ex}$ and
$\Delta_B$.

We expect that the  strongest variations for an ensemble of
(In,Al)As/AlAs QDs occur in the exciton radiative lifetimes,
$\tau_R$. This parameter is determined mainly by the $\Gamma-X$
mixing of the electronic states at the heterointerface. The smoother
the interface, the weaker the mixing~\cite{Shamirzaev84}. The
lifetime distribution is broad even in a given energy
window~\cite{Shamirzaev84,Rautert,Debus,Ivanov97}, which is
evidenced by a power law decay of the PL, see Fig.~\ref{fig2}.

It is difficult to accurately account for the spread of the QD
parameters in the theory. However, the effecte of longitudinal
(Faraday geometry) and transverse (Voigt geometry) magnetic fields
on the dynamics of the unpolarized PL, Fig.~\ref{fig2},
qualitatively agree with the theoretical predictions for a
moderately strong exchange interaction shown in Fig.~\ref{fig11}(b).

To describe the experiments below, we consider two sets
of identical QDs. These QDs represent the cases of fast and slow
radiative recombination. The QDs of the first set are described by
the following parameters: $\tau_R = 0.27 \, \mu {\rm s}$, $\tau_s^e
= 20 \, \mu {\rm s}$, $\tau_{NR} = 150\,\mu{\rm s} $,
$\tau_2=10^3\,\mu s$, $\tau_s^h=\tau_1=\tau_{NR} = \infty$,
$\Delta_B = 4.7\, {\rm mT}$, and $B_{ex} = 5 \,{\rm mT}$. The second set
of QDs have the same parameters except for $\tau_R = 24\, \mu{\rm
s}$ and $\tau_s^e = \infty$. The number of the former QDs is $1.7$
times larger than that of the latter.

\begin{figure}[t]
\centering
\includegraphics[width=0.95\linewidth]{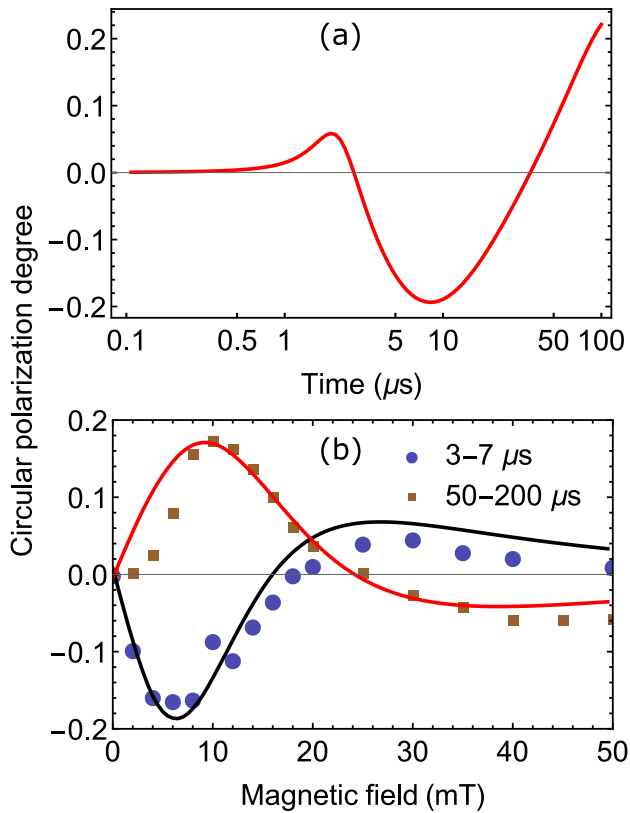}
\caption{(a) Dynamics of the PL polarization degree calculated for
an external magnetic field of $10$~mT accounting for the additional
unpolarized contribution, as described in the text. (b) Comparison
of experimental data (symbols) from Figs.~\ref{fig6}(a)
and~\ref{fig6}(b) with theoretical results (lines) for the
parameters given in  text. } \label{fig16}
\end{figure}

The dispersion of the exciton lifetimes is most important for the
dynamics of the PL polarization shown in Fig.~\ref{fig5}. However,
it can also be qualitatively reproduced theoretically, as we
demonstrate in Fig.~\ref{fig16}(a). Here we additionally accounted
for the unpolarized contribution to the PL, which decays with the
time constant of $0.27~\mu$s. In terms of physics, this contribution
can be related, for example, with the direct band gap QDs in the
ensemble.

The measurement of the PL at a particular time reduces the effect of
the lifetime distribution, as the QDs with the same average lifetime
mostly contribute to the PL at this time. The comparison between
experiment and theory is presented in Fig.~\ref{fig16}(b), where the
PL polarization degree is shown as a function of the external
magnetic field for two time intervals. The overall agreement is
quite good, and, in particular, the change of the sign of
polarization observed in experiment is reproduced theoretically.

Finally, we note that some features of the polarization dependence
on the magnetic field tilt angle shown in Fig.~\ref{fig7}(a) are
also captured by the theoretical model. Namely, small tilt angles
hardly affect the polarization in small magnetic fields and enhance
the polarization in large fields, as shown theoretically in
Fig.~\ref{fig15}. However, for a strong deviation from the Faraday
geometry the experimental data shown in Fig.~\ref{fig7}(b) are very
different from the theoretical predictions shown in the inset of
Fig.~\ref{fig15}. This is a challenge for our future investigations.

\section{Conclusion}
\label{sec:conclusions}

The circular polarization of the PL from indirect band gap
(In,Al)As/AlAs QDs has been studied experimentally in magnetic
fields of different orientations up to a few hundreds of millitesla
field strength, using nonresonant and unpolarized laser excitation.
The polarization of PL appears as result of the dynamic electron
spin polarization. We have shown that the PL polarization degree can
change sign up to two times, depending on the time delay after the
excitation pulse and the field strength and orientation. Most of the
experimental findings have been explained using a theoretical model
accounting for the electron-hole exchange interaction and the
hyperfine interaction. The dispersion of the QD parameters plays a
significant role in the calculation of the PL polarization degree,
open questions still remain with respect to the experimental results
in magnetic fields strongly tilted from the structure growth axis.

{\bf Acknowledgements.} We thank
\href{https://orcid.org/0000-0003-4462-0749}{M. M. Glazov} and
\href{https://orcid.org/0000-0002-0454-342X}{M. O. Nestoklon} for
fruitful discussions. The experimental part of this research has been
supported by the Deutsche Forschungsgemeinschaft (Grant
No.~409810106) and by the Russian Foundation for Basic Research
(Grants Nos.~19-52-12001 and 19-02-00098). M.B. acknowledges the
support by the Deutsche Forschungsgemeinschaft (ICRC TRR 160,
project A01). The theoretical studies by D.S.S. were supported by
the RF President Grant No. MK-5158.2021.1.2, the Foundation for the
Advancement of Theoretical Physics and Mathematics ``BASIS'', and
the Russian Foundation for Basic Research Grants Nos.~19-52-12054
and 20-32-70048. The theoretical studies by A.V.S. were supported by
the Russian Foundation for Basic Research Grant No.~19-02-00184.

% \cDY{Dima S., please check to whom relate these thanks}.
% The partial financial support by the RF President Grant No. MK-5158.2021.1.2 and the Foundation for the Advancement of Theoretical Physics and Mathematics ``BASIS''.

%\commentDima{Concerning RFBR Grants, mine are 19-52-12054 and 20-32-70048. Andrei acknowledges Grant 19-02-00184, and Timur --- 19-02-00098. Dima (Y.) do you acknowledge RFBR Grant 19-52-12001 and Deutsche Forschungsgemeinschaft via the project No.~409810106?}% \commentDima{I'm totally fine with the present wording, but some }


\begin{thebibliography}{}
\bibitem{Apex} The preliminary empirical tight binding calculations suggest that for small QDs the electron ground state might belong to the $X_z$ valley and be localized at the apex of the QDs~\cite{Nestoklon} similarly to (In,Ga)As/GaP QDs~\cite{Cedric1,Cedric2}.
\bibitem{relations} These notations are related to the notations of Ref.~\cite{Smirnov125} as follows: $N^{(\pm)}=N/2\pm J_z/3$, $\bm S^{(\pm)}=\bm S/2\pm\bm Q/2$.
\end{thebibliography}
\end{document}